\documentclass{llncs}
\usepackage{makeidx}  %
\usepackage{url}
\usepackage[boxruled,vlined]{algorithm2e}

\begin{document}

\title{Lightweight LCP Construction \\ for Next-Generation Sequencing Datasets}
\author{Markus J. Bauer\inst{1}, Anthony J. Cox\inst{1},\\
Giovanna Rosone\inst{2} and Marinella Sciortino\inst{2}}
\institute{Illumina Cambridge Ltd., United Kingdom\\
\email{\{mbauer,acox\}@illumina.com}
\and
University of Palermo, Dipartimento di Matematica e Informatica, Italy\\
\email{\{giovanna,mari\}@math.unipa.it}
}

\maketitle              %

\begin{abstract}

The advent of ``next-generation" DNA sequencing (NGS) technologies  has meant that collections of hundreds of millions of DNA sequences are now commonplace in bioinformatics. Knowing the longest common prefix array (LCP) of such a collection would facilitate the rapid computation of maximal exact matches, shortest unique substrings and shortest absent words.
CPU-efficient algorithms for computing the LCP of a string have been described in the literature, but require the presence in RAM of large data structures. This prevents such methods from being feasible for NGS datasets.

In this paper we propose the first lightweight method that simultaneously computes, via sequential scans, the LCP and BWT of very large collections of sequences. Computational results on collections as large as 800 million 100-mers demonstrate that our algorithm scales to the vast sequence collections encountered in human whole genome sequencing experiments.
\end{abstract}

\def\bwtS#1{\textsf{bwt}_{#1}(\mathcal{S})}
\def\lcpS#1{\textsf{lcp}_{#1}(\mathcal{S})} %
\def\bwtw#1{\textsf{bwt}_{#1}({w})}
\def\lcpw#1{\textsf{LCP}_{#1}({w})}
\def\bigS{\mathcal{S}}
\def\BCR{\texttt{BCR}}
\def\BWT{BWT} %
\def\LCP{LCP} %
\def\SA{SA} %
\def\GSA{GSA} %
\def\BCRLCP{\texttt{extLCP}}
\def\EMBWT{\texttt{BCRext} }
\def\EMBWTpp{\texttt{BCRext++} }
\def\BWTE{\texttt{bwte}}
\def\sort#1{\textrm{sort}(#1)}

\newcommand{\fbwt}{{\mathcal B}{\mathcal W}{\mathcal T}}

\section{Introduction}

The longest common prefix array (\LCP) of a string contains the lengths of the longest common prefixes of the suffixes pointed to by adjacent elements of the suffix array (\SA) of the string \cite{PuglisiTurpin2008}. The most immediate utility of the \LCP\ is to speed up suffix array algorithms and to simulate the more powerful, but more resource consuming, suffix tree. When combined with the suffix array or the Burrows-Wheeler transform (\BWT) of a string the \LCP\ facilitates, among other things, the rapid search for maximal exact matches, shortest unique substrings and shortest absent words \cite{OhlebuschGogKugelSpire2010,BellerGogOhlebuschSchnattinger2011,Herold_Kurtz_Gieg2008,Abouelhoda2004}. 
Existing algorithms for computing the \LCP\ require data structures of size proportional to the input data to be held in RAM, which has made it impractical to compute the \LCP\ of massive datasets such as the collections of hundreds of millions of reads produced by so-called Next-Generation Sequencing (NGS) technologies.

In this context, the aim of our paper is designing an algorithm for the computation of the \LCP\ of large collections of strings which works on an external memory system, by performing disk data accesses only via sequential scans, and is lightweight in the sense that its working space requirement is very low. 

Computing the \LCP\ of a collection of strings has been considered in the literature \cite{Shi:1996}. Defining $N$ and $K$ as respectively the sum of the lengths of all strings and the length of the longest string in the collection, the described approach requires $O(N\log K)$ time, but the $O(N \log N)$ bits of internal memory needed to store the collection and its \SA\ in internal memory prevents the method from scaling to massive data.

One can note that several algorithms to compute the \LCP\ of a single string in semi-external memory (see for instance \cite{KarkkainenManziniPuglisi:2009}) or directly via \BWT\ (see \cite{BellerGogOhlebuschSchnattinger2011}) could be adapted to solve the problem of computing the \LCP\ of a collection of strings. It could be sufficient to concatenate all the members of the collection into a single string and use distinct end marker symbols as separators.
However, assigning a different end marker to each string is not feasible when the collection is very large, but the alternative of terminating each member with the same symbol could lead to \LCP\ values that exceed the lengths of the strings and that depend on the order in which the strings are concatenated. In our approach, we compute the \LCP\ of the collection directly from the strings, without needing to concatenate them and without requiring precomputed auxiliary information such as the \SA\ or \BWT\ of the collection.
In fact, building upon the method of \BWT\ computation introduced in \cite{BauerCoxRosoneCPM11}, our algorithm adds some lightweight data structures and allows the \LCP\ and \BWT\ of a collection of $m$ strings to be computed simultaneously in $O((m+\sigma^2) \log(N))$ bits of memory, with a worst-case time complexity of $O(K(m+\sort{m}))$, where $\sort{m}$ is the time taken to sort $m$ integers, $\sigma$ is the size of the alphabet, $N$ is the sum of the lengths of all strings and $K$ is the length of the longest string.

The low memory requirement enables our algorithm to scale to the size of dataset encountered in human whole genome sequencing datasets: in our experiments, we compute the \BWT\ and \LCP\ of collections as large as $800$ millions $100$-mers.

Section~\ref{sec:prel} gives preliminaries that we will use throughout the paper, whereas Section~\ref{sec:algorithm} describes the sequential computation of the \LCP. We present details on the efficient implementation of the algorithm and computational results on real data in Sections~\ref{sec:implementation} and \ref{sec:experiments}, respectively.

\section{Preliminaries}
\label{sec:prel}

Let $\Sigma =\{c_1, c_2, \ldots, c_\sigma\}$ be a finite ordered alphabet with $c_1< c_2< \ldots < c_\sigma$, where $<$ denotes the standard lexicographic order.
We append to a finite string $w\in \Sigma^*$ an end marker symbol $\$$ that satisfies $\$ < c_1$. We denote its characters by $w[0], w[1],\ldots,w[k]$, where $k+1$ is the \emph{length} of $w$, denoted by $|w|$.
Note that, for $i<k$, $w[i]\in \Sigma$ and $w[k]=\$$.
A \emph{substring} of a string $w$ is written as $w[i,j] = w[i] \cdots w[j]$, with a substring $w[0,j]$ being called a \emph{prefix}, while a substring $w[i,k]$ is referred to as a \emph{suffix}.

The \emph{suffix array} of a string $w$ is an array $SA$ containing the permutation of the integers $0\ldots |w|-1$ that arranges the starting positions of the suffixes of $w$ into lexicographical order.
There exist some natural extensions of the suffix array to a collection of sequences (see \cite{Shi:1996}).
We denote by $\bigS$ the collection of $m$ strings $\{w_0,w_1,\ldots,w_{m-1}\}$.  %
We append to each sequence $w_i$ an end marker symbol $\$_i$ smaller than $c_1$, and $\$_i<\$_j$ if $i<j$. 
Let us denote by $N$ the sum of the lengths of all strings in $\bigS$. %

Let us denote by $\bigS_{(\mathrm{Pos},\mathrm{Seq})}$ the suffix starting at the position $\mathrm{Pos}$ of the string $w_{\mathrm{Seq}}$. We define the \emph{generalized suffix array} $GSA$ of the collection $\bigS$ as the array of $N$ pairs $(\mathrm{Pos},\mathrm{Seq})$, sorted by the lexicographic order of their corresponding suffixes $\bigS_{(\mathrm{Pos},\mathrm{Seq})}$. In particular, $GSA[i]=(t,j)$ is the pair corresponding to the $i$-th smallest suffix of the strings in $\bigS$. %

The \emph{longest common prefix array} (denoted by \LCP) of a collection $\bigS$ of strings is an array storing the length of the longest common prefixes between two consecutive suffixes of $\bigS$ in the lexicographic order. For every $j=1, \ldots, N-1$, if $GSA[j-1]=(p_1,p_2)$ and $GSA[j]=(q_1,q_2)$, $LCP[j]$ is the length of the longest common prefix of suffixes starting at positions $p_1$ and $q_1$ of the words $w_{p_2}$ and $w_{q_2}$, respectively. We set $LCP[0]=0$.

Note that the generalization of the suffix array to a collection $\bigS$ of strings is related to an extension of the notion of the \emph{Burrows-Wheeler transform} to a collection of strings that is a reversible transformation introduced in \cite{MantaciRRS07} (see also \cite{MantaciRRS08}). Actually, in its original definition, such a transformation produces a string that is a permutation of the characters of all strings in $\bigS$ but it does not make use of any end marker. 

In this paper we suppose that a different end marker is appended to each string of $\bigS$. Let us denote by $BWT(\bigS)$ the Burrows-Wheeler transform of the collection $\bigS$ and its output is produced according to the generalized suffix array of $\bigS$. In particular, if $GSA[i]=(t,j)$ then $BWT[i] = w_j[(t-1) \mathrm{mod} |w_j|]$.

Note that the output of $BWT(\bigS)$ differs, for at least $m$ symbols, from $BWT$ applied to the string obtained by concatenating all strings in $\bigS$. 
External memory methods for computing $BWT(\bigS)$ are given in \cite{BauerCoxRosoneCPM11}.

\section{\LCP\ computation of a collection of strings via \BWT}
\label{sec:algorithm}

The main goal of this section is to describe the strategy to compute the \LCP\ of a massive collection of strings via sequential scans of the disk data. In particular, the main theorem of the section enables the simultaneous computation of both \LCP\ and \BWT\ of a collection $\bigS$ of $m$ strings $\{w_0,w_1,\ldots,w_{m-1}\}$. We suppose that the last symbol of each sequence $w_i$ is the end marker $\$_i$. 
Our method scans all the strings from right to left and both \LCP\ and \BWT\ are incrementally built by simulating, step by step, the insertion of all suffixes having the same length in the generalized suffix array.

We refer to the suffix starting at the position $|w_i|-j-1$ of a string $w_i$ as its \emph{$j$-suffix}.
With the end marker $\$_i$ included, the $j$-suffix is of length $j+1$; the \emph{0-suffix} contains $\$_i$ alone.
Let us denote by $\bigS_j$ the collection of the $j$-suffixes of all the strings of $\bigS$. %

Let us denote by $K$ the maximal length of the strings in $\bigS$ and by $\lcpS{j}$ the longest common prefix array of the collection $\bigS_j$. It is easy to see that when $j=K$, $\lcpS{j}$ coincides with the \LCP\ of $\bigS$.
Since all $m$ end-markers are distinct, the longest common prefix of any pair of the $0$-suffixes is $0$, so the first $m$ positions into $\lcpS{j}$ are $0$ for any $j\geq 0$.

Note that $\lcpS{j}$ can be considered to be the concatenation of $\sigma +1$ arrays $L_j(0),L_j(1), \ldots,L_j(\sigma)$ where, for $h=1,\ldots,\sigma$, the array $L_j(h)$ is the \LCP\ of the suffixes of $\bigS_j$ that start with  $c_h \in \Sigma$, while $L_j(0)$ (corresponding to the $0$-suffixes) is an array of $m$ zeroes.
It is easy to see that $\lcpS{0}=L_0(0)$ and that $L_0(h)$ is empty for $h>0$. For sake of simplicity, each segment is indexed starting from $1$. We note that, for each $0<h\leq\sigma$, $L_j(h)[1]=0$ and $L_j(h)[i]\geq 1$ for $i>1$.

Similarly, we define the string $\bwtS{j}$ as the Burrows-Wheeler transform of the collection of the $j$-suffixes of $\bigS$. This can be partitioned in an analogous way into segments $B_j(0),B_j(1), \ldots,B_j(\sigma)$, where the symbols in $B_j(0)$ are the characters preceding the lexicographically sorted $0$-suffixes of $\bigS_j$ and the symbols in $B_j(h)$, with $h \geq 1$, are the characters preceding the lexicographically sorted suffixes of $\bigS_j$ starting with $c_h \in \Sigma$. Moreover, $\bwtS{0}=B_0(0)$ and the segments $B_0(h)$ are empty for $h>0$.

In this section we show that, for each $j > 0$, $\lcpS{j}$ can be sequentially constructed by using $\bwtS{j-1}$ and $\lcpS{j-1}$ (in previous work \cite{BauerCoxRosoneCPM11}, three of the present authors showed how $\bwtS{j}$ may be computed from $\bwtS{j-1}$).
Note that $\bwtS{0}$ and $\lcpS{0}$ are defined above.

Given the segments $B_j(h)$ and $L_j(h)$, $h=0,\ldots,\sigma$, for the symbol $x$ occurring at position $r$ of $B_j(h)$ we define the $(j,h)$\emph{-LCP Current Interval} of $x$ in $r$ (denoted by $LCI_j^h(x,r)$) as the range $(d_1,r]$ in $L_{j}(h)$, where
$d_1$ is the greatest position smaller than $r$ of the symbol $x$ in $B_{j}(h)$, if such a position exists. If such a position does not exist, we define $LCI_j^h(x,r)=L_j(h)[r]$.
Analogously, we define for the symbol $x$ the $(j,h)$\emph{-LCP Successive Interval} of $x$ in $r$ (denoted by $LSI_j^h(x,r)$) as the range $(r,d_2]$ in $L_{j}(h)$, where
$d_2$ is the smallest position greater than $r$ of the symbol $x$ in $B_{j}(h)$, if it exists. If such a position does not exist we define $LSI_j^h(x,r)=L_j(h)[r]$. In our notation, a range is delimited by a square bracket if the correspondent endpoint is included, whereas the parenthesis means that the endpoint of the range is excluded.

Actually, it is easy to verify that $d_1=\mathsf{select}(\mathsf{rank}(x,r)-1,x)$ and $d_2=\mathsf{select}(\mathsf{rank}(x,r)+1,x)$, where $\mathsf{rank}(x,r)$ counts the number of $x$'s until position $r$ and $\mathsf{select}(p,x)$ finds the position of the $p$-th occurrence of $x$ in a segment $B_{j}$.

The following theorem shows how to compute the segments $L_j(h)$, with $j > 0$, by using $L_{j-1}(h)$ and $B_{j-1}(h)$ for any $h > 0$.
We denote by $\textrm{Suf}_j(0)$ the lexicographically sorted $0$-suffixes and by $\textrm{Suf}_j(h)$, for $h>0$, the lexicographically sorted $t$-suffixes of $\bigS_j$, with $t\leq j$ starting with $c_h \in \Sigma$.

\begin{theorem}\label{th:LCP_case_allsuffix}
Let $\mathcal{I}=\{r_0 < r_1< \ldots< r_{q-1}\}$ be the set of the positions in $\textrm{Suf}_j(z)$ of the $j$-suffixes starting with the letter $c_z$.
For each position $r_p \in \mathcal{I}$ ($0 \leq p < q$),
$$L_j(z)[r_p]=\left\{\begin{array}{ll}
                     0 & \mbox{ if $r_p=1$} \\
                     1 & \mbox{ if $r_p>1$ and $LCI_{j-1}^v(c_z,t)=L_{j-1}(v)[t]$} \\
                     \min LCI_{j-1}^v(c_z,t) +1& \mbox{otherwise}
                   \end{array}\right.
                   $$
where $c_v$ is the first character of the $(j-1)$-suffix of $w_i$, and $t$ is the position in $B_{j-1}(v)$ of symbol $c_z$ preceding the $(j-1)$-suffix of $w_i$.\\

For each position $(r_p+1) \notin \mathcal{I}$ (where $r_p \in \mathcal{I}$ and $0 \leq p < q$), then
$$L_j(z)[r_p+1]=\left\{\begin{array}{ll}
                     1 & \mbox{ if $LSI_{j-1}^v(c_z,t)=L_{j-1}(v)[t]$} \\
                     \min LSI_{j-1}^v(c_z,t) +1& \mbox{otherwise}
                   \end{array}\right.
                   $$

For each position $s$, where $1 \leq s<r_p$ (for $p=0$), $r_{p-1} < s < r_p$ (for $0 < p < q-1$), $s> r_p$ (for $p=q-1$) then
$$L_j(z)[s]=L_j(z)[s-p]$$
\end{theorem}

For lack of space, the proof of the theorem is omitted and we defer it in the full paper.

A consequence of the theorem is that the segments $B_{j}$ and $L_{j}$ can be constructed sequentially and stored in external files.
This fact will be used in the next section.

\section{Lightweight implementation via sequential scans}\label{sec:implementation}

Based on the strategy described in the previous section, here we propose an algorithm (named \BCRLCP) that simultaneously computes the \BWT\ and \LCP\ of a set of strings $\bigS$. Memory use is minimized by reading data sequentially from files held in external memory: only a small proportion of the symbols of $\bigS$ need to be held in RAM and we do not need to keep the generalized suffix array of $\bigS$. Obviously, the generalized suffix array can be a side output of our implementation. 

Our method extends previous work \cite{BauerCoxRosoneCPM11,BauerCoxRosoneTCS2012} on computing the \BWT\ of a collection of strings and we follow the notation therein.

Although our algorithm is not restricted to collections of strings of uniform length, for sake of simplicity our description supposes that $\bigS$ comprises $m$ strings of length $k$ and we assume that $j = 0,1,\ldots,k$ and $i=0,1,\ldots,m-1$. We simulate $m$ distinct end-markers by using a single end-marker $\$=c_0$ and setting $w_s[k] < w_t[k]$ if and only if $s < t$, so that if two strings $w_s$ and $w_t$ share the $j$-suffix, then $w_s[k-j,k] < w_t[k-j,k]$ if and only if $s < t$. Moreover, we assume that the values of $\lcpS{j}$ do not exceed $j$ and the first $m$ positions into $\lcpS{j}$ are $0$ for any $j\geq 0$.
The main part of the algorithm consists of $k$ consecutive iterations. At iteration $j$, we consider all the $j$-suffixes of $\bigS$ and simulate their insertion in the GSA.
In other words, for each $i$, we have to find the position of the suffix $w_i[k-j,k]$ according to the lexicographic order of all the suffixes of $\bigS$ of length at most $j$, then insert the new symbol circularly preceding the $j$-suffix of $w_i$ into $B_{j}(z)$, where $c_z=w_i[k-j]$, for some $z=1,\ldots,\sigma$, and update the values in $L_{j}(z)$.
Consequently, both $\bwtS{j}$ and $\lcpS{j}$ are updated accordingly.
Note that, at each iteration $j$, both the segments $B_{j}$ and $L_{j}$, initially empty, are stored in different external files that replace the files used in the previous iteration.

In order to compute $\bwtS{j}$ and $\lcpS{j}$, the algorithm needs to hold six arrays of $m$ integers in internal memory. Four of these ($P_j$, $Q_j$, $N_j$ and $U_j$) are useful to compute the \BWT\  (see \cite{BauerCoxRosoneCPM11}), a further two ($C_j$ and $S_j$) are needed to compute and update the values of the longest common prefixes. we will give a description of all these arrays but, for brevity, we will focus on the computation of $C_j$ and $S_j$.

Each of the arrays $P_j$, $Q_j$, $N_j$ and $U_j$ contains $m$ elements, as detailed in the following. At the end of iteration $j$, if $w_i[k-j,k]$ is the $q$-th $j$-suffix then:
\begin{itemize}
\item $N_j[q]$ contains the index $i$. It uses $O(m \log m)$ bits of workspace.
\item $Q_j[q]$ stores the index $z$ where $c_z=w_i[k-j]$, i.e. the first symbol of the $j$-suffix. It uses $O(m \log \sigma)$ bits of workspace.
\item $P_j[q]$ contains the position in $B_j(z)$ of the symbol circularly preceding the $j$-suffix $w_i[k-j,k]$, such a symbol is $w_i[k-j-1]$ and it is stored at the position $N_j[q]$ of $U_j$. So, it needs $O(m \log (mk))$ bits of workspace.
\item $U_j$ stores the new characters to be inserted, one for each sequence in $\bigS$, so it uses $O(m \log \sigma)$ bits of workspace.
\end{itemize}

The arrays $C_j$ and $S_j$ each contain $m$ integers useful to compute $\lcpS{j}$ for $j>0$.  In particular, $C_j[q]$ stores the value in \LCP\ between the $j$-suffix $w_i[k-j,k]$ and the previous suffix in the GSA with respect to the lexicographic order of all the suffixes of $\bigS$ of length at most $j$, whereas $S_j[q]$ contains the value in \LCP\ between the $j$-suffix and the next suffix in GSA (if it exists). Such values will be computed at the iteration $j-1$ according to Theorem \ref{th:LCP_case_allsuffix}.
We observe that $C_j$ and $S_j$ contain exactly one integer for each sequence in the collection and they use $O(m \log k)$ bits of workspace.

At the iteration $j=0$, the algorithm initializes the segments $B_0$ and $L_0$ as described in previous section, i.e. $B_0(0)=w_0[k-1]w_1[k-1]\cdots w_{m-1}[k-1]$ and for each $q=0,\ldots,m-1$, we set $L_{0}(0)[q]=0$. Consequently, for $q=0,\ldots,m-1$, the arrays are initialized by setting $N_{0}[q]=q$, $P_{0}[q]=q+1$, $Q_{0}[q]=0$, $C_{1}[q]=1$ and  $S_{1}[q]=1$.

For I/O efficiency, each iteration $j>0$ can be divided into two consecutive phases: during the first one we read only the segments $B_{j-1}$ in order to find the arrays $P_j$, $Q_j$ $N_j$ and $U_j$. In phase 2, the segments $B_{j-1}$ and $L_{j-1}$ are read once sequentially both for the construction of new segments $B_{j}$ and $L_{j}$ and for the computation of the arrays $C_{j+1}$ and $S_{j+1}$, as they will be used in the next iteration.
In the following we describe both the phases of the generic iteration $j>0$.
Figure \ref{fig:compute} illustrates the execution of the algorithm for a simple collection at the iterations $12$ and $13$. 
In the first phase the arrays $P_j$, $Q_j$ and $N_j$ are computed. In particular, if $w_i[k-j-1]$ (or the end marker $\$$ for the last step) is the new symbol to be inserted, its position $r$ is obtained by computing the number of occurrences of $c_z=w_i[k-j]$ in $B_{j-1}(0),\ldots, B_{j-1}(v-1)$ and in $B_{j-1}(v)[1,t]$, where $c_v=w_i[k-(j-1)]$ and $t$ is the position of $c_z$ in $B_{j-1}(v)$. Hence, the index $z$ is stored into $Q_j$ at some position $q$, the computed position $r$ (where storing the new symbol) is added to the array $P_j[q]$ and $i$ is added to $N_j[q]$.
Note that in order to find the positions, a table of $O(\sigma^2 \log (mk))$ bits of memory is used.
Finally we sort $Q_{j}$, $P_{j}$, $N_j$, $C_{j}$, $S_j$ where the first and the second keys of the sorting are the values in $Q_j$ and $P_j$ respectively.

\begin{figure}[!htb]
{\scriptsize
  $$
    \begin{array}{cccl}
  &  L_{12}(0) &  B_{12}(0) & \mbox{Sorted Suffixes} \\
  & 0 & C &       \$_0 \\
  & 0 & C &       \$_1 \\
  &  L_{12}(1) &  B_{12}(1) & \mbox{Sorted Suffixes} \\
\multicolumn{ 1}{c|}{} & 0 & G & AAAGCTC\$_1 \\
\cline{2-2}
\multicolumn{ 1}{c|}{LCI_{12}(G,3)} & 2 & C &    AAC\$_0 \\
\multicolumn{ 1}{c|}{\rightarrow} &    {\bf 3} &    {\bf G} & {\bf AACAGAAAGCTC\$_1} \\
  & 2 & A & AAGCTC\$_1 \\
  & 1 & A &     AC\$_0 \\
  & 2 & A & ACAGAAAGCTC\$_1 \\
  & 2 & T & ACCAAC\$_0 \\
  & 2 & C & ACTGTACCAAC\$_0 \\
  & 1 & C & AGAAAGCTC\$_1 \\
  & 2 & A &  AGCTC\$_1 \\
  &   &   &   \\
  &  L_{12}(2) &  B_{12}(2) & \mbox{Sorted Suffixes} \\
\multicolumn{ 1}{c|}{} & 0 & A &      C\$_0 \\
\cline{2-2}
\multicolumn{ 1}{c|}{LCI_{12}(A,4)} & 1 & T &      C\$_1 \\
\multicolumn{ 1}{c|}{} & 1 & C &   CAAC\$_0 \\
\multicolumn{ 1}{c|}{\rightarrow} &    {\bf 2} &    {\bf A} & {\bf CACTGTACCAAC\$_0} \\
\cline{2-2}
\multicolumn{ 1}{c|}{LSI_{12}(A,4)} & 2 & A & CAGAAAGCTC\$_1 \\
  & 1 & A &  CCAAC\$_0 \\
  & 1 & G &    CTC\$_1 \\
  & 2 & A & CTGTACCAAC\$_0 \\
  &  L_{12}(3) &  B_{12}(3) & \mbox{Sorted Suffixes} \\
  & 0 & A & GAAAGCTC\$_1 \\
  & 1 & A &   GCTC\$_1 \\
  & 1 & T & GTACCAAC\$_0 \\
  &   &   &   \\
  &  L_{12}(4) &  B_{12}(4) & \mbox{Sorted Suffixes} \\
  & 0 & G & TACCAAC\$_0 \\
  & 1 & C &     TC\$_1 \\
  & 1 & C & TGTACCAAC\$_0 \\
    \end{array}%
\qquad
    \begin{array}{cccl}
    & L_{13}(0) &  B_{13}(0) & \mbox{Sorted Suffixes} \\
    &     0 & C &       \$_0 \\
    &     0 & C &       \$_1 \\
 & L_{13}(1) &  B_{13}(1) & \mbox{Sorted Suffixes} \\
   &      0 & G & AAAGCTC\$_1 \\
   &      2 & C &    AAC\$_0 \\
   &      3 & G & AACAGAAAGCTC\$_1 \\
   &      2 & A & AAGCTC\$_1 \\
   &      1 & A &     AC\$_0 \\
   \rightarrow & {\bf 2} & {\bf \$_0} & {\bf ACACTGTACCAAC\$_0} \\
   &      \textbf{\underline{3}} & A & ACAGAAAGCTC\$_1 \\
   &      2 & T & ACCAAC\$_0 \\
   &      2 & C & ACTGTACCAAC\$_0 \\
   &      1 & C & AGAAAGCTC\$_1 \\
   &      2 & A &  AGCTC\$_1 \\
 & L_{13}(2) &  B_{13}(2) & \mbox{Sorted Suffixes} \\
   &      0 & A &      C\$_0 \\
   &      1 & T &      C\$_1 \\
   &      1 & C &   CAAC\$_0 \\
   &      2 & G & CACTGTACCAAC\$_0 \\
   &      2 & A & CAGAAAGCTC\$_1 \\
   &      1 & A &  CCAAC\$_0 \\
   &      1 & G &    CTC\$_1 \\
   &      2 & A & CTGTACCAAC\$_0 \\
 & L_{13}(3) &  B_{13}(3) & \mbox{Sorted Suffixes} \\
   &     0 & A & GAAAGCTC\$_1 \\
  \rightarrow & {\bf 3} & {\bf \$_1} & {\bf GAACAGAAAGCTC\$_1} \\
   &      \textbf{\underline{1}} & A &   GCTC\$_1 \\
   &      1 & T & GTACCAAC\$_0 \\
& L_{13}(4) &  B_{13}(4) & \mbox{Sorted Suffixes} \\
   &        0 & G & TACCAAC\$_0 \\
   &      1 & C &     TC\$_1 \\
   &      1 & C & TGTACCAAC\$_0 \\
     \end{array}%
   $$
}\caption{
Iteration $12$ (on the left) and iteration $13$ (on the right) on the collection $\bigS=\{ACACTGTACCAAC,GAACAGAAAGCTC\}$. We append different end-marker to each string ($\$_0$ and $\$_1$, respectively) to make the explanation more immediate but the same situation would occur using the same symbol. The first two columns represent the partial \LCP\ and the partial \BWT\ after the iterations. The positions of the new symbols corresponding to the $13$-suffixes (shown in bold on the right) are computed from the positions of the $12$-suffixes (in bold on the left), which were retained in the array $P$ after the iteration $12$. The new values in \LCP\ (shown in bold on the right) are computed during the iteration $12$ and are contained in $C_{12}$. The updated values in \LCP\ (shown in bold and underlined on the right) are computed during the iteration $12$ and are contained in $S_{12}$.
}\label{fig:compute}
\end{figure}

Here we focus on the second phase in which the computation of the segments $L_j$ is performed by using the arrays $C_{j}$ and $S_{j}$ constructed during the previous step.
Note that the sorting of the arrays allows us to open and sequentially read the pair files ($B_{j-1}(h)$ and $L_{j-1}(h)$ for $h=0,\ldots,\sigma$) at most once.

For all symbols in $U_j$ that we have to insert in the segment $B_j(h)$, the crucial point is to compute $C_{j+1}$ by using $LCI_{j}^h$ and $S_{j+1}$ by using $LSI_{j}^h$  while the new files are being constructed, instead of using auxiliary data structures to compute \textsf{rank} and \textsf{select}.
For each index $z$, we consider all the elements in $Q_j$ equal to $z$. Because of the sorting, such elements are consecutive.
Let $0\leq l,l'\leq m-1$ be their first and the last positions, respectively. Hence for each $l \leq p \leq l'$, we have $Q_j[p]=z$ and $P_{j}[l]<  \ldots < P_{j}[l']$.
In order to apply Theorem \ref{th:LCP_case_allsuffix}, we need to compute $LCI_{j}^h$ and $LSI_{j}^h$ of each new symbol in its new position.
Since each $B_j(h)$ and $L_j(h)$ are constructed sequentially, we do not know \emph{a priori} the opening positions $LCI_{j}^h$ and the closing positions $LSI_{j}^h$ that are used to compute $C_{j+1}$ and $S_{j+1}$.
However we can observe that when we write a symbol $x$ into $B_j(h)$, its occurrence could be the opening or closing positions of some $LCI_{j}^h$ and $LSI_{j}^h$ of $x$, if $x$ is a new symbol. Such considerations are outlined in detail in the following.

For each symbol $\alpha$ that we insert at position $s$ in $B_j(z)$, with $1 \leq s<P_{j}[l]$,  it is easy to see that $B_j(z)[s]=B_{j-1}(z)[s]$ and $L_j(z)[s]=L_{j-1}(z)[s]$. Moreover, the position of $\alpha$ could be the opening position of $LCI_{j}^z(\alpha, y)$, if $\alpha$ is the new symbol that will be inserted at some next position $y$.

For each new symbol $\beta$ that we insert at position $P_{j}[q]$ in $B_j(z)$ ($l\leq q\leq l'$), we have $\beta=U_j[N_{j}[q]]$ and, by Theorem \ref{th:LCP_case_allsuffix}, it follows that  $L_j(z)[P_{j}[q]]=0$ if $P_{j}[q]=1$ or $L_j(z)[P_{j}[q]]=C_j[q]$ otherwise. Moreover:
\begin{itemize}
    		\item The position $P_j[q]$ surely is the closing position of $LCI_{j}^z(\beta, P_j[q])$. If the position $P_j[q]$ is the first occurrence of $\beta$ in $B_{j}(z)$, then $LCI_{j}^z(\beta, P_j[q])=L_j(z)[P_j[q]]$ and we set $C_{j+1}[q]=L_j(z)[P_j[q]]+1$ according to Theorem \ref{th:LCP_case_allsuffix}. Otherwise, we set $C_{j+1}[q] = \min(LCI_{j}^z(\beta, P_j[q]))+1$, whose computation has been started when the interval was opened.

    \item The position $P_j[q]$ could be the opening position of $LCI_{j}^z(\beta, y)$, if $\beta$ will be inserted, as new symbol, at some next position $y$.

    \item The position $P_j[q]$ could be the closing position of $LSI_{j}^z(\beta, y)$, where $y$ represents, eventually, the largest position $P_j[f]$, with $P_j[f]<P_j[q]$, for $l \leq f < q$, where $\beta$ has been inserted. In this case, we set $S_{j+1}[f] = \min (LSI_{j}^z(\beta, P_j[q]))+1$ in according with Theorem \ref{th:LCP_case_allsuffix}.

	\item The position $P_j[q]$ surely is the opening position of $LSI_{j}^z(\beta, P_j[q])$.
We observe that if the position $P_j[q]$ is the last occurrence of $\beta$ in $B_{j}(z)$ (we will discover this at the end of the file), it means that $LSI_{j}^z(\beta, P_j[q])=L_j(z)[P_j[q]]$, i.e. $S_{j+1}[q]=1$. 				
\end{itemize}
			
For each symbol $\alpha$ that we insert at position $(P_{j}[q]+1)$, with $P_{j}[q]+1 \neq P_{j}[q+1]$,  $B_j(z)[P_{j}[q]+1]=B_{j-1}(z)[P_{j}[q-p]]$ (where $p$ is the number of the new symbols already inserted) and, by Theorem \ref{th:LCP_case_allsuffix}, $L_j(z)[P_{j}[q]+1]=S_j[q]$.
	For each symbol $\alpha$ that we insert at position $s$ in $B_j(z)$, with $P_{j}[q] < s < P_{j}[q+1]$ ($l < q \leq l'$), we have $B_j(z)[s]=B_j(z)[s-p]$ and, by Theorem \ref{th:LCP_case_allsuffix},
$L_j(z)[s]=L_{j-1}(z)[s-p]$, where $p$ is the number of the new symbols already inserted. Moreover:
\begin{itemize}
\item The position $s$ could be the opening position of $LCI_{j}^z(\alpha, y)$, if $\alpha$ will be inserted, as new symbol, at some next position $y$.
\item The position $s$ could be the closing position of $LSI_{j}^z(\alpha, P_j[f])$, if $\alpha$ has been inserted, as new symbol, at some previous position $P_j[f]$, with $P_j[f]<P_j[q]$, for $l \leq f < q$. In this case, we set $S_{j+1}[f] = \min (LSI_{j}^z(\alpha, P_j[f]))+1$ according to Theorem~\ref{th:LCP_case_allsuffix}.
    \end{itemize}	
For each symbol $\alpha$ that we insert at the position $s$ in $B_j(z)$, where $s > P_{j}[l']$, we have $B_j(z)[s]=B_j(z)[s-(l'-l+1)]$ and, by Theorem \ref{th:LCP_case_allsuffix}, 			 $L_j(z)[s]=L_{j-1}(z)[s-(l'-l+1)]$. Moreover, the position of $\alpha$ could be the closing position of $LSI_{j}^z(\alpha, P_j[f])$, if $\alpha$ has been inserted as a new symbol at some position $P_j[f]$, for $l \leq f \leq l'$. In this case, we set $S_{j+1}[f]=\min(LSI_{j}^z(\alpha, P_j[f]))+1$ in according with Theorem~\ref{th:LCP_case_allsuffix}.

When $B_j(z)$ is entirely built, the closing position of some $LSI_{j}^z(\alpha, y)$ could remain not found. This means that the last occurrence of $\alpha$ appears at position $y$. Note that $y$ must be equal to some $P_j[f]$, $l \leq f \leq l'$. In this case, we set $S_{j+1}[f]=1$ according to Theorem~\ref{th:LCP_case_allsuffix}.

It is easy to verify that we can run these steps in a sequential way. Moreover, one can deduce that, while the same segment is considered, for each symbol $\alpha\in \Sigma$ at most one $LCI_{j}^h(\alpha,t)$ for some $t$, and at most one $LSI_{j}^h(\alpha,r)$ for some $r\leq t$, will have not their closing position. For this reason we use two arrays $minLCI$ and $minLSI$ of $\sigma$ integers that store, for each symbol $\alpha$ in $\Sigma$, the minimum among the values of \LCP\ in the possible corresponding $LCI$ or $LSI$ without closing position, respectively.

From the size of the data structures and from the above description of the phases of the \BCRLCP\ algorithm, we can state the following theorem.

\begin{theorem} 
Given a collection $\bigS$ of $m$ strings of length $k$ over an alphabet of size $\sigma$, the \emph{\BCRLCP} algorithm computes  \BWT\ and \LCP\ of $\bigS$ by using $O(mk^2 \log \sigma)$ disk I/O and $O((m+\sigma^2) \log (mk))$ bits of memory in $O(k(m+\sort{m})$ CPU time, where $\sort{m}$ is the time taken to sort $m$ integers.
\end{theorem}

The following corollary describes the performance of the method when the collection contains strings of different length.

\begin{corollary}
Given a collection $\bigS$ of $m$ strings over an alphabet of size $\sigma$, the \LCP\ and \BWT\ of $\bigS$ are computed simultaneously in $O((m+\sigma^2) \log(N))$ bits of memory, with a worst-case time complexity of $O(K(m+\sort{m}))$, where $\sort{m}$ is the time taken to sort $m$ integers, $N$ is the sum of the lengths of all strings and $K$ is the length of the longest string.
\end{corollary}

\section{Computational experiments and discussion}
\label{sec:experiments}

To assess the performance of our algorithm on real data, we used a publicly available collection of human genome sequences from the Sequence Read Archive \cite{SRA11} at \url{ftp://ftp.sra.ebi.ac.uk/vol1/ERA015/ERA015743/srf/} and created subsets containing $43$, $85$, $100$, $200$ and $800$ million reads, each read being 100 bases in length.
We developed \BCRLCP, an implementation of the algorithm described in Section~\ref{sec:implementation}, which is available upon request from the authors. 
Our primary goal was to analyze the additional overhead in runtime and memory consumption of simultaneously computing both \BWT\ and \LCP\ via \BCRLCP\ compared with the cost of using \BCR\ (\cite{BauerCoxRosoneCPM11}) to compute the \BWT\ alone.

Table~\ref{tab:inputInstances} shows the results for the instances that we created. We do see increase in runtime since
 \BCRLCP\ writes the values of \LCP\ after that the symbols in \BWT\ are written, so it effectively increases the I/O operations. 
So, a time optimization could be obtained if we read/write at the same time both the elements in \BWT\ and \LCP\ by using two different disks. 
All tests except the $800$ million read instance were done on the same machine, having $16$Gb of memory and two quad-core Intel Xeon E5450 $3.0$GHz processors. Although the \LCP\ of a collection of $700$ million $100$-mers was successfully computed on the same machine using $15$Gb of RAM, the collection of $800$ million reads needed slightly more than $16$Gb so was processed on a machine with $64$Gb of RAM and four quad-core Intel Xeon E7330 $2.4$GHz processors. On both machines, only a single core was used for the computation.
Moreover, to examine the behaviour of our algorithm on reads longer than $100$bp, we created a set of $50$ million $200$bp long reads based on the $100$ million $100$bp instance. It turns out that, although the sheer data volume is the same, \BCRLCP\ uses $1.2$Gb and takes $10.3$ microseconds per input base.

Our algorithm represents the first lightweight method that simultaneously computes, via sequential scans, the \LCP\ and \BWT\ of a vast collection of sequences. Recall that the problem of the \LCP\ computation of a collection of strings has been faced in \cite{Shi:1996}, but such a strategy works in internal memory. 
Recently, however, some lightweight approaches for the \LCP\ computation of a single string were described in the literature.
Some of them use of the suffix array of the string \cite{KarkkainenManziniPuglisi:2009,Fischer2011,PuglisiTurpin2008,GogOhlebuschO11},
but the space needed to hold this in RAM is prohibitive for NGS datasets. 
However, in \cite{BellerGogOhlebuschSchnattinger2011}, the authors give an algorithm for the construction of the \LCP\  of a string that acts directly on the BWT of the string and does not need its suffix array.
A memory-optimized version of this algorithm \cite{BellerGogOhlebuschSchnattinger2012} (called {\tt{bwt\_based\_laca2}}) needs to hold the \BWT\ of the string in internal memory plus a further $1.5n$ bytes, where $n$ is the length of the input string.
\begin{table}[!htb]
\begin{center}
    \begin{tabular}{p{1.5cm}rp{.2cm}p{2cm}p{2cm}p{2cm}p{2cm}}
     \hline \hline%
     \textbf{instance } & \textbf{size} & & \textbf{program} & \textbf{wall clock} & \textbf{efficiency} & \textbf{memory}\\
                        \hline \hline%
   0043M & $4.00$  & & \BCR\        & $0.99$          & $0.84$ & $0.57$\\ %
         & $4.00$  & & \BCRLCP        & $3.29$          & $0.98$ & $1.00$\\
\hline
   0085M &        $8.00$  & & \BCR\        & $1.01$       & $0.83$ & $1.10$ \\ %
         &        $8.00$  & & \BCRLCP     & $3.81$       & $0.87$ & $2.00$ \\
\hline
   0100M & $9.31$  & & \BCR\    & $1.05$            & $0.81$ & $1.35$\\   %
        & $9.31$  & & \BCRLCP & $4.03$            & $0.83$ & $2.30$\\ %
\hline
   0200M & $18.62$  & & \BCR\ & $1.63$            & $0.58$ & $4.00$ \\ %
         & $18.62$  & & \BCRLCP & $4.28$             & $0.79$ & $4.70$ \\ %
\hline
\hline
   0800M & $74.51$ & & \BCR\ & $3.23$            & $0.43$ & $10.40$\\ %
   		   & $74.51$ & & \BCRLCP & $6.68$           &  $0.67$ &  $18.00$\\  %
\hline
\end{tabular}
\end{center}
\caption{The input string collections were generated on an Illumina GAIIx sequencer, all reads are $100$ bases long.
Size is the input size in gigabytes, wall clock time---the amount of time that elapsed
from the start to the completion of the instance---is given as microseconds per
input base, and memory denotes the maximal amount of memory (in gigabytes) used
during execution. The efficiency column states the CPU efficiency values,
i.e. the proportion of time for which the CPU was occupied and not waiting
for I/O operations to finish, as taken from the output of the
\texttt{/usr/bin/time} command. %
}
\label{tab:inputInstances}
\end{table}

Notice that an entirely like-for-like comparison between our implementation and the above existing implementation for \BWT\ and \LCP\ computation of a string would imply the concatenation of the strings of the collection by different end markers. However, for our knowledge, the existing implementations do not support the many millions of distinct end markers our test collections would require. 

An alternative is to concatenate each of strings with the same end marker. This leads to values in the \LCP\ that may possibly exhibit the undesirable properties of exceeding the lengths of the strings and depending on the order in which the strings are concatenated, but does allow the \BWT\ of the resulting string to be computed in external memory by using the algorithm \BWTE\ proposed in \cite{Ferragina10}.

The combined \BWT/\LCP\  computation provided by \BCRLCP\ has a faster runtime than \BWTE. In particular, for the $0085M$ instance, \BWTE\ uses $14$Gb of memory and needs $3.84$ microseconds per input base vs \BCRLCP\ that uses $2$Gb of memory and $3.81$ microseconds per input base.

We have also used \BCR\, by suitable preprocessing steps, to simulate the computation of the \BWT\ of the concatenated strings.
We compared \BCR, \BCRLCP\ and {\tt{bwt\_based\_laca2}} on the $0200M$ instance.
Since the memory consumption of {\tt{bwt\_based\_laca2}} exceeded $16$Gb on this dataset, we ran the tests on a machine of identical CPU to the $16$Gb machine, but with 64Gb RAM.

With \BCR, the \BWT\ was created in under $5$ hours of wallclock time taking only $4$Gb of RAM, while {\tt{bwt\_based\_laca2}} required $18$Gb of RAM to create the \LCP\ in about $1$ hour $45$ minutes. Our new method \BCRLCP\ needed $4.7$Gb of RAM to create both \BWT\ and \LCP\ in just under $18$ hours. Attempting to use {\tt{bwt\_based\_laca2}} to compute the \LCP\ of the $0800M$ instance exceeded the available RAM on the $64$Gb RAM machine.

The experimental results show that our algorithm is a competitive tool for the lightweight simultaneous computation of \LCP\  and \BWT\ on the string collections produced by NGS technologies.
Actually, the \LCP\  and \BWT\ are two of the three data structures needed to build a compressed suffix tree (CST) \cite{Sadakane2007} of a string. 
The strategy proposed in this paper could enable the lightweight construction of CSTs of strings collections for comparing, indexing and assembling vast datasets of sequences when memory is the main bottleneck.
Our current prototype can be further optimized in terms of memory by performing the sorting step in external memory. Further saving of the working space could be obtained if we embody our strategy in \EMBWT or \EMBWTpp (see \cite{BauerCoxRosoneTCS2012}). These methods, although slower than \BCR, need to store only a constant and (for the DNA alphabet) negligibly small number of integers in RAM regardless of the size of the input data.
\bibliographystyle{plain}	%
\bibliography{BWT}		%

\end{document}